\documentstyle[12pt,aaspp4]{article}

\lefthead{Bono et al.}

\begin{document}

\title{On the primordial helium content: CMB and stellar constraints}  

\author{
G. Bono\altaffilmark{1}, 
A. Balbi\altaffilmark{2},
S. Cassisi\altaffilmark{3}, 
N. Vittorio\altaffilmark{2}, and 
R. Buonanno\altaffilmark{1,2}, 
}

\affil{1. Osservatorio Astronomico di Roma, Via Frascati 33, 00040 Monte 
Porzio Catone, Italy; bono@mporzio.astro.it; buonanno@mporzio.astro.it}
\affil{2. Dipartimento di Fisica, Universit\`a Tor Vergata, and INFN, 
Sezione di Roma II, \\ Via della Ricerca Scientifica 1, 
00133 Roma, Italy; \\ balbi@roma2.infn.it; Nicola.Vittorio@roma2.infn.it}
\affil{3. Osservatorio Astronomico di Collurania, via M. Maggini,
64100 Teramo; cassisi@te.astro.it}

\begin{abstract}
We present the results of a joint investigation aimed at constraining 
the primordial He content ($Y_P$) on the basis of both the Cosmic 
Microwave Background (CMB) anisotropy 
and two stellar observables, namely the tip of the Red Giant Branch (TRGB) 
and the luminosity of the Zero Age Horizontal Branch (ZAHB). 
Current baryon density estimates based on CMB measurements cover a
wide range values $0.009\la \Omega_bh^2 \la 0.045$, that according to 
Big Bang Nucleosynthesis (BBN) models would imply $0.24\la Y_P \la 0.26$. 
We constructed several sets of evolutionary tracks and HB models by 
adopting $Y_P=0.26$ and several metal contents.
The comparison between theory and observations suggests that ZAHB magnitudes
based on He-enhanced models are 1.5$\sigma$ brighter than the empirical ones.
The same outcome applies for the TRGB bolometric magnitudes. This finding
somewhat supports a $Y_P$ abundance close to the canonical 0.23-0.24 value.
More quantitative 
constraints on this parameter are hampered by the fact that the CMB 
pattern shows a sizable dependence on both $Y_P$ and the baryon 
density only at small angular scales, i.e. at high $l$ in the power 
spectrum ($l\ga 100$). However, this region of the power spectrum 
could be still affected by deceptive systematic uncertainties.  

Finally, we suggest to use the {\em UV-upturn} to estimate the He
content on Gpc scales. In fact, we find that a strong increase in $Y_P$
causes in metal-poor, hot HB structures a decrease in the UV emission.
\end{abstract}


\pagebreak 
\section{Introduction} 

The comparison between chemical abundances of deuterium, helium, and 
lithium predicted by BBN models with current empirical 
estimates is one of the most viable method to constrain the physical 
mechanisms and the cosmology which governed the nucleosynthesis of 
primordial abundances (Olive, Steigman, \& Walker 2000).  

As far as the primordial He content is concerned, current empirical 
estimates are mainly based on measurements of nebular emission lines 
in low-metallicity, extragalactic HII regions (Izotov, Thuan, \&
Lipovetsky 1997; Olive, Steigman, \& Skillman 1997). Recent 
He determinations present small observational errors ($\approx 1$\%), 
but large uncertainties between independent measurements:  
$Y_P=0.234\pm0.003$ by Olive \& Steigman (1995) against 
$Y_P=0.244\pm0.002$ by Izotov \& Thuan (1998). This evidence suggests 
that current He abundances are still dominated by systematic errors. 
In fact, Viegas, Gruenwald, \& Steigman (2000) and Gruenwald, Steigman, 
\& Viegas (2001) in two detailed 
investigations on the ionization correction for unseen neutral and 
doubly-ionized He in HII regions, found that He estimates should be 
reduced by 0.006 ($Y_P=0.238\pm0.003$), a quantity which is a factor 
of 2-3 larger than typical statistical errors quoted in the literature.  
Moreover and even more importantly, Pistinner et al. (1999) on the 
basis of a new grid of stellar atmosphere models for OB stars 
found that the inclusion of both NLTE and metal-line blanketing 
effects causes an increase of the order of 40\% in the ratio of 
He to H ionizing photons. This evidence together with uncertainties 
due to the occurrence of stellar winds, shocks, temperature 
fluctuations (Izotov, Thuan, \& Lipovetsky 1997; Pistinner et al. 1999; 
Peimbert, Peimbert, \& Luridiana 2001; Sauer, \& Jedamzik 2001, and 
references therein) and of peculiar nebular dynamics certainly affects the 
He abundance estimates based on giant extragalactic HII regions.  
In addition it is worth mentioning that the HII regions used for determining
the cosmological Helium abundance could have been somewhat polluted by 
the stellar yields of the pristine type II Supernovae, and in turn the 
empirical He abundances in these stellar systems should be corrected 
for self-pollution by massive stars. A plain evidence of this 
occurrence has been recently provided by Aloisi, Tosi, \& Greggio (1999),  
and \"Ostlin (2000). On the basis of deep HST optical and NICMOS 
data they have resolved the stellar content of I~ZW~18 and found evidence 
that this blue compact galaxy hosts a relatively old population of 
asymptotic giant branch stars  ($\approx$ 0.1-5 Gyr).

On the other hand, the comparison between star counts of 
horizontal branch (HB, central He burning phase) and red giant 
(RG, H shell burning phase) stars in Galactic Globular Clusters 
(GGCs) with the lifetimes predicted by evolutionary models, 
the so-called R parameter (Iben 1968), supplies upper limits 
to primordial He mass fraction of the order of 0.20 
(Sandquist 2000; Zoccali et al. 2000). However, such estimates 
should be cautiously treated (Bono et al. 1995; Cassisi et al. 1998), 
since they are hampered by current uncertainties on the nuclear 
cross-section of the $^{12}C(\alpha,\gamma)^{16}O$ reaction 
(Buchmann 1996).  
Note that spectroscopic measurements of He abundances in low-mass 
population II stars are useless for constraining the primordial 
He content, because the He lines are either too faint 
(low-temperature stars) or affected by gravitational settling such 
as high temperature HB stars (Giannone \& Rossi 1981; 
Moheler et al. 1999).  

However, empirical and statistical errors affecting abundance 
determinations of primordial deuterium, $^3$He, and lithium could 
be significantly larger than for He 
(Sasselov \& Goldwirth 1995; Olive et al. 2000). Moreover, the 
primordial He content plays a paramount role in constraining 
both stellar ages and cosmic distances, since the Mass-Luminosity 
(M/L) relation of low and intermediate-mass stars during H and He 
burning phases depends on $Y_P$ (Bono et al. 2000). At the same time, 
at fixed He to metal enrichment ratio the He abundance adopted 
to model evolutionary and pulsational properties of metal-rich 
stellar structures does depend on $Y_P$ as well (Bono et al. 1997; 
Zoccali et al. 2000). 

The physical baryon density of the universe is one of the observables
that can be determined with high accuracy using measurements of CMB 
anisotropies at intermediate and small angular scales 
(see e.g., Hu et al. 2000, and references therein). 
It goes without saying that this observable 
plays a key role not only to assess the plausibility of the 
physical assumptions adopted in BBN models (Tegmark \& Zaldarriaga 
2000) but also for constraining the intrinsic accuracy of current 
primordial abundance estimates. 
According to the joint analysis of both BOOMERanG and MAXIMA-1 data,
it has been estimated at 68\% confidence level a baryon density  
$\Omega_B h^2= 0.032_{-0.004}^{+0.005}$ (Jaffe et al. 2001). 
On the basis of this observable 
Esposito et al. (2001) found that the new CMB measurements are 
inconsistent at more that 3$\sigma$ with both standard and degenerate 
BBN models. 

On the other hand, the latest analysis of the BOOMERanG results,
which has improved the removal of systematics from the data
(Netterfield et al. 2001), found $\Omega_b h^2=0.022^{+0.004}_{-0.003}$
(de Bernardis et al. 2001), in very good agreement with the BBN value.
The same conclusion has been derived from the analysis of the
ground-based CMB observations performed by the DASI interferometer
(Halverson et al. 2001), which also found
$\Omega_b h^2=0.022^{+0.004}_{-0.003}$ (Pryke et al. 2001).
This notwithstanding, the new analysis of the MAXIMA data (Lee et al. 2001),
which extended the high $l$ coverage of the power spectrum measurement,
still points towards somewhat higher values of the physical baryon
density: $\Omega_b h^2=0.032\pm 0.006$ (Stompor et al.  2001).
Finally, we mention the measurements of the CMB power spectrum
at $l > $1000 by the Cosmic Background Imager. From these observations,
and by assuming a flat cosmological model, the likelihood for the
physical baryon density is found to peak at $\Omega_b h^2$=0.009
(Padin et al. 2001). Current physical baryon densities based on CMB
measurements and BBN models would imply that $Y_P$ might range from 
roughly 0.24 to approximately 0.26.
 
The main aim of this investigation is to constrain $Y_P$ on the basis 
of recent CMB measurements and two stellar observables that depend on
$Y_P$, namely the TRGB luminosity and the ZAHB luminosity.
In \S 2 we discuss in detail the adopted theoretical framework
as well as the comparison between predicted and empirical observables.
The effect of a change in $Y_P$ abundance on the {\em UV-upturn}
as well as on CMB anisotropies are presented in \S 3 and \S 3.1
respectively.  Our conclusions and final remarks are briefly
mentioned in \S 4.

\section{Stellar constraints on $Y_P$}

On the basis of the new CMB measurements and BBN models Esposito et al. 
(2000) estimated at 68\% confidence level a $Y_P$ abundance ranging 
from 0.249 to 0.254. However, recent investigations (Tegmark \& 
Zaldarriaga 2000; Padin et al. 2001; Stompor et al. 2001; 
Tegmark, Zaldarriaga, \& Hamilton 2001) suggest 
on the basis of CMB measurements and of the simplest flat inflation 
model that the baryon density should range from 0.009 to 0.045. 
The upper limit taken at face value would imply 
a larger primordial He content. As a generous but still plausible 
primordial He content we adopted $Y_P=0.26$. To assess the impact 
that such a determination has on stellar structures we selected 
two observables, namely the luminosity of ZAHB stars and the 
luminosity of the tip of TRGB stars. 
The previous observables refer to stars belonging to GGCs. 
The reasons why we selected these observables are the following: 
1) the stellar population in GGCs are among the oldest stars in the 
Galaxy, and therefore they are the best laboratory to investigate 
the primordial He content. The comparison between theory and observations 
in GGCs is more predictable when compared with field, halo stars, since it 
relies on stars that are coeval, located at the same distance, and chemically
homogeneous\footnote{Note that the initial He content in stellar populations 
typical of GGCs is an upper limit for $Y_P$, since the latter was 
somewhat contaminated by the debris of the first stellar generation.}. 
2) The ZAHB in GGCs marks the phase in which the stars are mainly 
supported by $3\alpha$ reaction in the stellar center and it is a well-defined 
observational feature. The ZAHB luminosity depends on the He core mass 
and an increase of 15\% in Y causes an increase in the luminosity of 
approximately 0.16 mag (Sweigart \& Gross 1976; Raffelt 1990).  
3) According to current evolutionary prescriptions the TRGB phase marks 
the onset of central He burning (He-core flash) in low-mass stars and 
its luminosity strongly depends on the He-core mass, and in turn on the 
initial He content. In fact, Sweigart \& Gross (1978) found that an 
increase of 15\% in 
Y causes a decrease in the TRGB luminosity of 0.1 mag. Finally, 
we mention that the previous observables are virtually unaffected 
by cluster age ($\approx 13\pm3$ Gyr; Vandenberg, Stetson, \& Bolte 1996), 
since for stellar ages larger than $\approx$6 Gyr both the ZAHB and the TRGB 
luminosities do not depend on age (Castellani, Degl'Innocenti, \& 
Luridiana 1993; Lee, Freedman, \& Madore 1993; Cassisi \& Salaris 1997; 
Salaris \& Cassisi 1998).

Fig. 1 shows the comparison between predicted and observed ZAHB luminosity
at the RR Lyrae effective temperature ($\log T_e=3.85$) as a function of 
global metallicity\footnote{The global metallicity is a parameter which 
accounts for both iron and $\alpha-element$ abundances (Salaris et al. 1993;
Vandenberg et al. 2000).}. The approach adopted to estimate the empirical 
bolometric magnitudes as well as their errors have been discussed by  
De Santis \& Cassisi (1999).  
The solid and the dashed line show the ZAHB luminosities predicted 
by Cassisi \& Salaris (1997) and by Vandenberg et al. (2000). The comparison 
between theory and observations suggests that HB models constructed by 
the previous authors are in good agreement with empirical data. On the 
other hand, the ZAHB luminosities predicted by HB models constructed by 
assuming $Y_P=0.26$ (dotted line) are $1.5\sigma$ brighter than the observed 
ones. Note that current predictions on ZAHB luminosities are still 
controversial, since there is a mounting evidence that recent HB models 
based on new input physics (equation of state, neutrino energy loss rates, 
conductive opacities) are systematically brighter than observed and 
predicted by canonical HB models  (Cassisi et al. 1999; 
Bono, Castellani, \& Marconi 2000). 
However, such a conundrum does not affect our conclusion, because the 
systematic shift in the luminosity showed by He-enhanced HB models 
is a differential effect. 
 
Fig. 2 shows the comparison between theory and observations for the 
TRGB bolometric magnitudes in a sample of 12 GGCs as a function of 
global metallicity. Apparent bolometric magnitudes were estimated by 
Frogel, Persson, \& Cohen (1983) and by Ferraro et al. (2000). The distance 
moduli adopted to derive the absolute magnitudes were estimated by comparing 
predicted ZAHB luminosities at fixed chemical composition, derived by adopting 
the same theoretical framework, with the observed distribution of HB stars 
in the Color-Magnitude diagram of each individual cluster.  
Theoretical predictions (solid line) refer to models constructed by adopting 
a canonical initial He content of 0.23 and global metallicities ranging from 
[M/H]=-2.4 to -0.4. The top panel shows that the predicted luminosities are 
systematically brighter than the observed ones. This mismatch between theory 
and observations is expected and caused by the 
fact that the empirical estimates are hampered by sample size and also 
by the decrease in the lifetimes of the stellar structures approaching 
the TRGB. By taking into account the typical sample sizes of cluster RGB 
stars, Salaris \& Cassisi (1997) found that the 30\% of clusters 
should lay within 0.1 mag (dotted line) the predicted TRGB luminosity 
and the 70\% within 0.3 mag (dashed line) the predicted ones. Data plotted 
in the top panel support, within current uncertainties, this prediction. 
As a matter of fact, more than 70\% of the empirical TRGB bolometric 
magnitudes lay within the expected range (dashed and dotted lines) and 
only a few clusters attain magnitudes close to predicted $M^{tip}_{Bol}$
values.  

The bottom panel of Fig. 2 shows the same empirical data plotted in the 
top panel. Theoretical predictions on TRGB bolometric magnitudes are based 
on evolutionary models constructed by adopting an initial He content of 0.26. 
At the same time, the absolute bolometric magnitudes were estimated by 
adopting the distance moduli obtained by comparing observed HB stars with 
the ZAHB luminosities predicted by HB models constructed by adopting 
$Y_P=0.26$, i.e. the same He content adopted in He enhanced models of 
Fig. 1. A glance at the data plotted in this panel shows that more 
than 70\% of the clusters in our sample attain bolometric magnitudes 
similar or even brighter than predicted by He enhanced models.  
This finding is at odds with the straightforward statistical arguments
mentioned above, and indeed only two measurements lay within 0.3 mag 
from the predicted TRGB bolometric magnitudes.

\section{The dependence of the {\em UV-upturn} on $Y_P$}

Stellar observables discussed in the previous section can be extended
over Local Group (LG) galaxies or slightly beyond, i.e. on scales of the
order of a few Mpc. Moreover, current He estimates are based on
spectroscopic measurements of HII regions in extremely metal-poor blue
compact galaxies. These systems are located outside the LG and their
typical distances are, within the uncertainties, of the order of
200-300 Mpc. This means that previous stellar observables and HII regions
can hardly be adopted on Gpc scale even by using the largest telescopes.
However, the He abundance on these scales can be probed on the basis of 
the {\em UV-upturn}. 
This phenomenon shows up as a sharp rise in the spectra of elliptical 
and S0 galaxies for wavelengths smaller than 2500 $\AA$. According to 
both theoretical (Greggio \& Renzini 1990; Castellani \& Tornamb\`e 1991;
Castellani et al. 1994) and empirical (Burstein et al. 1988; 
Brown et al. 2000) evidence the {\em UV-upturn} is driven by the progeny 
of extreme and hot HB stars, namely AGB-manqu\`e and post-early-AGB 
(Castellani \& Tornamb\`e 1991). This observable presents a strong 
dependence on age, since He abundance affects the mass of main sequence 
turn off stars,  the HB morphology, and in turn the UV emission of old 
stellar populations. 
As a consequence, the {\em UV-upturn} is a powerful age indicator in 
elliptical galaxies (Tantalo et al. 1996; Greggio \& Renzini 1999).   

However, as already mentioned in the previous section the evolutionary 
properties of HB stars do depend on the primordial He content. 
In particular, an increase in Y causes, at fixed age, an increase 
in the ZAHB luminosity\footnote{Note that the ZAHB luminosity 
is governed by the size of the He-core and by He abundance in the envelope.
However, the ZAHB luminosity of hot HB stars mainly depends on the 
He-core mass due to the marginal efficiency of the H-burning shell. 
Therefore, an increase in the initial He abundance causes a decrease in 
the He-core mass at the He-flash, and in turn a decrease in the luminosity 
of hot HB stars.}. Current HB evolutionary models for metal-poor 
and metal-rich structures ($0.0001 \le Z \le 0.02$) constructed 
(Zoccali et al. 2000) by adopting different He contents 
($0.20 \le Y_P \le 0.26$) suggest that the HB luminosity at $\log T_e=3.85$ 
scales with the initial He content according to the following relation: 
${\Delta \log L_{HB}/\Delta Y}\approx 1.8$. A quite similar value was 
also suggested by Raffelt (1990) on the basis of HB models computed 
by Sweigart \& Gross (1976, 1978). 
Thus suggesting that evolutionary predictions on this ratio are quite 
robust. This means that in metal-rich populations, which is typical 
of E and S0 galaxies, an increase in Y of a factor of two (0.23 vs 0.46) 
causes an increase in $\log L_{HB}$ of the order of 0.40 dex.  

This result is a rough estimate of the impact of Y on the HB luminosity,
it cannot be easily extrapolated to the {\em UV-upturn}. In fact, 
the crucial parameters for this phenomenon are the changes in HB and 
post-HB evolutionary lifetimes as well as the UV flux and not the 
bolometric one. This means that a detailed quantitative 
estimate does require synthetic models which simultaneously account 
for both H and He burning phases, and in turn for the change in the 
spectral energy distribution (SED) of the entire field population
(Brown et al. 1998). 
This notwithstanding, we are interested in estimating the dependence 
of the {\em UV-upturn} on He content at intermediate redshifts, and 
therefore we constructed four metal-poor (Z=0.0001) extreme HB structures 
at different He contents, namely 0.15, 0.23, 0.35, and 0.5 (see Fig. 3). 
The He-core masses of these structures were estimated by constructing 
H-burning evolutionary tracks that reach the RGB tip with an age of 
$\approx$ 10 Gyr. We find that the He-core masses for the previous 
compositions are 0.523, 0.511, 0.481, and 0.433 $M_\odot$ respectively. 
To estimate the impact of He content on the UV emission of AGB-manqu\`e 
stars we selected a typical effective temperature  
for these structures, namely $\log T_e\approx4.45$. One finds that 
the total mass of HB structures at the selected compositions range 
from $M/M_\odot=$ 0.53 ($Y_P=0.15$) to 0.45 ($Y_P=0.50$). 
The evolution was followed from central He-burning till the beginning of 
the white dwarf cooling sequence (see Fig. 3), and then we estimated the 
total spectral energy distribution (SED) for the four structures according 
to the individual evolutionary lifetimes. It turns out that the UV emission 
of Extreme Horizontal Branch (EHB) structures is relatively sensitive 
to the He content, and indeed an 
increase from $Y_P=0.23$ to $Y_P=0.35$ and $Y_P=0.50$ causes a decrease in 
the UV emission by roughly 10\% and 22\% respectively. At the same time, 
a decrease from $Y_P=0.23$ to $Y_P=0.15$ causes an increase by $\approx$9\%. 
This effect is due to the fact that among EHB structures an increase in 
Y causes, at fixed age and effective temperature, a decrease in the 
He-core mass and in turn in the ZAHB luminosity. This means that He-rich 
EHB structures spend a substantial portion of their central He-burning 
lifetime at lower luminosities. Therefore their total UV emission 
decreases when compared with He-poor structures.  
Moreover, we also find that an increase in the He content causes 
in metal-poor structures a decrease in the range of stellar masses 
evolving at high temperatures during He-burning. In fact, we find 
that the largest mass that evolve as AGB-manqu\`e slightly decreases 
from 0.54 at $Y_P=0.15$ to 0.52 at $Y_P=0.35$. This finding further 
strengthens the evidence that in metal-poor structure an increase 
in the He abundance 
causes a decrease in the UV emission. Note that such a trend is at 
odds with the behavior in metal-rich structure, and indeed Dorman 
et al. (1993) found that an increase in Y causes an increase in the 
largest mass evolving as AGB-manqu\`e, and in turn in the 
UV emission. The difference seems to be due to the fact that the 
evolution of metal-poor EHB structures is mainly governed by central 
He burning, and therefore by the He-core mass at the tip of the RGB. 
On the contrary the H-burning shell is more efficient in metal-rich 
EHB structures. This means that the He-core mass in these structures 
undergoes a mild increase during the central He burning phase, and therefore 
the range of stellar masses evolving as AGB-manqu\`e increases as well. 
Once again we note that current arguments are preliminary but plausible 
speculations of the impact of He-content on the {\em UV-upturn}. 
However, a firm quantitative evaluation requires the calculation of 
synthetic population models that account for both AGB-Manqu\`e and 
post-early-AGB structures. 

This seems a promising result, since a substantial increase/decrease in 
the He content should cause, at fixed look-back time and similar star 
formation histories, a decrease/increase in the scatter of the empirical 
average restframe 1550-V colors, i.e. the fingerprint of the {\em UV-upturn}. 
Moreover and even more importantly, present-day instrumentation allowed the 
detection and measurement of the {\em UV-upturn} in intermediate redshift 
($z\approx0.6$) E galaxies (Brown et al. 2000b). By assuming a Hubble 
constant $H_0=67$ km s$^{-1}$ Mpc$^{-1}$ and an Eistein-de Sitter cosmological 
model one finds that the comoving distance $D_M$ and the look-back time 
$t_l$ of this E galaxy are $\approx1.8$ Gpc and $\approx5$ Gyr respectively. 
On the other hand, if we assume a high lambda cosmological model 
($\Omega_M=0.2$ $\Omega_\lambda=0.8$) one finds $D_M\approx2.2$ Gpc 
and $t_l\approx6$ Gyr. This is the reason why we constructed EHB structures 
by adopting He-core masses at evolutionary ages of approximately 10 Gyr. 
Therefore this evidence and our finding seem to support the use of the 
{\em UV-upturn} to trace the He content up to Gpc scales.  

\subsection{The dependence of the CMB on $Y_P$}

As it has been emphasized countless times in the literature (see e.g., 
Jungman et al. 1996; Hu \& White 1996; Kamionkowski \& Kosowsky 1999), 
observations of CMB anisotropies provide a powerful way of setting
tight constraints on the value of most cosmological parameters. 
In particular, the angular power spectrum of CMB temperature 
fluctuations depends both on the primordial fluctuations which seeded 
structure formation in the universe and on the physical processes occurring
before the recombination in the baryon-photon plasma. Within the 
framework of inflationary adiabatic models, these processes leave a
characteristic imprint in the angular power spectrum in the form of a 
series of harmonic peaks (usually named ``acoustic peaks'' in the literature) 
whose height and position is sensitively dependent on the parameters of 
the cosmological model. 

While the parameter which is most robustly determined from measurements
of the CMB power spectrum is undoubtedly the total energy density of the 
universe, many other parameters are measurable with striking precision. 
Among them, the physical density of baryonic matter in the universe,
whose value affects the height ratio of odd and even peaks in the 
spectrum: in particular, a high baryon density enhances the height 
of the first peak with respect to the second, and vice-versa. This
effect is quite relevant even for small variations of $\Omega_b h^2$, as
shown in Fig. 4. The dependence of the CMB anisotropy on the primordial
He mass fraction is, on the contrary, quite weak. As shown in Figure
5, the effect on the first peak is just about $3\%$, even for an 
unrealistically large range of values $0.15\la Y_P \la 0.50$. 
One has to go beyond the second peak
in order to obtain effects larger than $10\%$ .
We stress the fact that the effect of both the baryon density and the 
primordial He abundance on the CMB pattern is only relevant at small 
angular scales (corresponding to $l\ga 100$ in the power spectrum).
These scales are only weakly affected by the uncertainty deriving 
from the limited coverage of the observed region, and from the so 
called ``cosmic variance'' resulting from the fact 
that we can only observe one statistical realization (our sky) drawn from
the underlying cosmological model. On the other hand, the high $l$ end of the 
spectrum is unfortunately the one which is currently most at risk of 
being affected by unknown systematics such as pointing inaccuracies, 
poorly known beam pattern, residual instrumental noise, etc. 

Until recent times, no high resolution observation of the CMB
anisotropy pattern was available. Consequently, the structure of peaks
in the power spectrum could not be resolved with the accuracy needed
to obtain a precise measurement of the baryon density.  The situation
has dramatically changed after recent observations. The MAXIMA and
BOOMERanG experiments (Hanany et al. 2000; de Bernardis et al. 2000)
produced the first high-resolution maps of the CMB, and measured the
CMB angular power spectrum on a wide range of $l$: $36\la l\la 1200$,
corresponding to angular scales $10\arcmin\la\theta\la 10\arcdeg$.
This has provided tight constraints on the main parameters of the
inflationary adiabatic model (Balbi et al. 2000; Lange et al. 2000;
Jaffe et al. 2001). In particular, the constraints on the physical
baryon density from the joint analysis of the MAXIMA and BOOMERanG
data, $\Omega_B h^2= 0.032_{-0.004}^{+0.005}$, was found to be higher
than the one derived from primordial nucleosynthesis considerations
$\Omega_b h^2=0.020\pm 0.002$ (see e.g., Burles, Nollett \& Turner
2001a; Burles et al. 1999) although the BBN value fall within the CMB
95\% confidence interval.  This stirred some discussion about the
existence of a conflict between CMB and BBN and possible explanation
for it (for example, Burles, Nollett \& Turner 2001b; Kurki-Suonio \&
Sihvola 2001; Esposito et al. 2001; Di Bari \& Foot 2001; Lesgourgues
\& Peloso 2000).  It is remarkable, however, that these first limits
from the CMB agree, within 2$\sigma$, with those from the BBN, which 
are derived using a different set of measurements.

\section{Summary and final remarks}

Recent measurements of the CMB anisotropy provided the unique
opportunity to evaluate several fundamental cosmological parameters
and to supply for all of them a preliminary but plausible estimate of
their error budget. The impact of these new measurements on cosmological
models uncorked a flourishing literature. However, it is not easy to
assess on a quantitative basis to what extent current differences in
the physical baryon density derived from CMB observations are caused
by deceptive systematic errors. As a matter of fact, CMB measurements
of $\Omega_bh^2$ range from 0.009 (Padin et al. 2001) to
$0.032\pm0.012$ (95\% confidence level, Stompor et al. 2001).
According to BBN models the new measurements imply that $Y_P$ might
range from approximately 0.24 to roughly 0.26. By adopting the upper 
limit on $Y_P$ we investigated the impact of the change on two stellar 
observables, 
namely the ZAHB luminosity and the luminosity of the tip of the RGB.
The main outcome of our analysis is that an increase in the primordial
He content from the canonical $Y_P=0.23$ to $Y_P=0.26$ does not seem to
be supported by the comparison between current theoretical predictions
and empirical data. 
 
We found that the {\em UV-upturn} can be adopted to estimate the 
primordial He content.  In fact, numerical
experiments suggest that an increase of $Y_P$ from 0.23 to 0.50 causes
a decrease in the UV emission at least of the order of 20\%.  This is
a preliminary rough estimate based on the assumption that AGB-manqu\`e
structures are the main sources of the {\em UV-upturn}.  An
interesting feature of this observable is that current instruments can
allow us to measure the {\em UV-upturn} up to distances of the order
of Gpcs. Note that to supply quantitative estimates of $Y_P$ on the
basis of the comparison between synthetic and observed {\em UV-upturns} 
it is necessary to account for the SED typical of
complex stellar populations as a function of redshift (Tantalo et al.
1996; Yi et al. 1999). However, theoretical predictions should be
cautiously treated, since UV flux when moving from low to high metal
contents strongly depends on the efficiency of the mass loss as well  
as on the He to metal enrichment ratio (Greggio \& Renzini 1999). The
scenario has been further complicated by recent spectroscopic
measurements of hot HB stars ($T_e \ge 10,000$ K) in metal-poor GGCs
(M15, M13).  In fact, Behr et al. (2000) and Behr, Cohen, \& McCarthy
(2000) found that in these stars the iron abundance is enhanced by 1-2
order of magnitudes, whereas the He content is depleted by at least
one order of magnitude respect to solar abundance.  Unfortunately, we
still lack quantitative estimates of the impact that such a
peculiarities have on the UV emission.

Theoretical and empirical arguments support the evidence that the
density of baryons in the universe is homogeneous (Copi, Olive, \&
Schramm 1995). The same outcome applies to large scale chemical
inhomogeneities (Copi, Olive, \& Schramm 1996). However, it has been
recently suggested by Dolgov \& Pagel (1999, hereinafter DP) a new
cosmological model that predicts a substantial spatial variation in the
primordial chemical composition and a small baryon density variation. This
investigation was triggered by a difference of one order of magnitude in
the deuterium abundance of damped $Ly\alpha$ systems along the line of
sight of high-redshift ($0.5 \le z \le 3.5$) QSOs (D'Odorico et al. 2001;
Steigman et al. 2001). The scenario  developed by DP relies on a model
of leptogenesis (Dolgov 1992) in which takes place a large lepton asymmetry
and this asymmetry undergoes strong changes on spatial scales ranging from
Mega to Giga pcs. The key feature of this model is to predict a large and
varying lepton asymmetry and a small baryon asymmetry. Within this
theoretical framework the He mass fraction in deuterium-rich regions
should range from 35\% to 60\%, while the $Li$ one should increase up
to $10^{-9}$, while the variation of the photon temperature should be
$\delta T/T\approx 2.5\times10^{-3}$. Obviously, the hypothesis that
current changes in baryon density are due to real spatial variations
is premature as any further speculative issue. 
Future full-sky CMB observations from space missions such as 
NASA's MAP (Wright 1999) and ESA's PLANCK (Mandolesi et al. 1998)
will play a crucial role to properly address the problem of the
spatial variation, since they will supply a larger sensitivity up 
to very high $l$ ($l > 1000$) and an improved control on systematics.

\acknowledgements 
We warmly thank Claudia Maraston for kindly providing us the spectral
energy distribution of current HB models. We also ackowledge an anonymous 
referee for his/her useful suggestions that improved the readability of 
the paper. This work was supported by MURST/Cofin2000 under the project: 
"Stellar Observables of Cosmological Relevance" (G.\ B., R.\ B., \& S.\ C.). 

\pagebreak

\pagebreak 

\figcaption{Bolometric magnitude of the Zero Age Horizontal Branch 
at the effective temperature of RR Lyrae stars ($\log T_e=3.85$) as 
a function of the global metallicity. Dotted and dashed lines show 
current theoretical predictions based on a primordial He content 
$Y_P=0.23$, while the solid line for $Y_P=0.26$. The latter is at 
odds with empirical estimates.}

\figcaption{Top panel: comparison between predicted (solid line) and 
empirical bolometric magnitude for the tip of the Red Giant Branch. 
To account for the typical sample size, the dotted and the dashed 
lines show the luminosity range within which should lay the 30\% 
and the 70\% of empirical tip luminosities. Squares refer to 
data collected by Frogel et al. (1983, filled) and by 
Ferraro et al. (2000, open). Bottom panel: same as the top panel 
but for $Y_P=0.26$.}

\figcaption{HR diagram for metal-poor (Z=0.0001) extreme HB models at 
different He contents, namely 0.15, 0.23, 0.35, and 0.50. The solid line 
shows the ZAHB for $Y_P=0.23$. The He-core masses were estimated by evolving 
H-burning structures that reach the RGB tip with an age of approximately 
10 Gyr. The total ZAHB masses (see labeled values) were selected to 
populate an effective temperature -$\log T_e\approx4.45$- typical of 
AGB-Manqu\`e structures. Empty circles mark steps in age of t=25 Myr 
during central He-burning, while diamonds mark steps in age of t=1.5 Myr 
during off-ZAHB evolution (double-shell burning).}   

\figcaption{The dependence of the CMB angular power spectrum on the 
physical baryon density $\Omega_b h^2$. The solid line was 
computed assuming a value $\Omega_b h^2=0.02$, as derived from BBN 
considerations.  The dashed line
assumes a slightly higher value $\Omega_b h^2=0.03$, as preferred
by the first analyses of the MAXIMA and BOOMERanG CMB data.
Both spectra were computed assuming an inflationary adiabatic model, with
scale invariant primordial perturbation, and with $\Omega=1$,  
$\Lambda=0.7$, $h=0.68$ and a primordial He mass fraction $Y_P=0.24$.}

\figcaption{The dependence of the CMB angular power spectrum on
the primordial He content. The solid line was computed
using a ``standard'' value of $Y_P=0.24$. The dashed line
assumes $Y_P=0.15$, while the dotted line has $Y_P=0.50$.
All spectra were computed assuming an inflationary adiabatic model, 
with scale invariant primordial perturbation, and with 
$\Omega=1$, $\Omega_b h^2=0.02$, $\Omega_{CDM} h^2=0.12$, 
$\Lambda=0.7$, $h=0.68$.}


\begin{references}
\reference{} Aloisi, A., Tosi, M., \& Greggio, L. 1999, AJ, 118, 302
\reference{} Balbi, A., et al., 2000, ApJL, 545, L1
\reference{} Behr, B. B., Cohen, J. G., McCarthy, J. K.  2000, ApJ, 531, L37
\reference{} Behr, B. B., Djorgovski, S. G., Cohen, J. G., 
McCarthy, J. K., Cot\`e, P., Piotto, G., Zoccali, M. 2000, ApJ, 528, 849
\reference{} Bono, G., Caputo, F., Cassisi, S., Incerpi, R., \& Marconi, M. 
1997, ApJ, 483, 811
\reference{} Bono, G., Caputo, F., Cassisi, S., Marconi, M., Piersanti, L., \&
Tornamb\'e, A. 2000, ApJ, 543, 955
\reference{} Bono, G., Castellani, V., Degl'Innocenti, S., \& Pulone, L. 1995, 
A\&A, 297, 115
\reference{} Bono, G., Castellani, V., \& Marconi, M. 2000, ApJ, 532, L129
\reference{} Brown, T. M., Bowers, C. W., Kimble, R. A., Sweigart, A. V., \& 
Ferguson, H. C. 2000, ApJ, 532, 308  
\reference{} Brown, T. M., Ferguson, H. C., Stanford, S. A., \& Deharveng, J.-M.
1998, ApJ, 504, 113  
\reference{} Burles, S., Nollett, K.M., \& Turner, M. 2001a, ApJL, 552, L1
\reference{} Burles, S., Nollett, K.M., \& Turner, M. 2001b, Phys. Rev. D, 63, 063512
\reference{} Burles, S., et al., 1999, Phys. Rev. Lett., 82, 4176
\reference{} Burstein, D., Bertola, F., Buson, L. M., Faber, S. M., 
\& Lauer, T. R. 1988, ApJ, 328, 440   
\reference{} Cassisi, S., Castellani, V., Degl'Innocenti, S., \& Weiss, A. 1998,
A\&AS, 129, 267
\reference{} Cassisi, S., Castellani, V., Degl'Innocenti, S., Salaris, M., \& 
Weiss, A. 1999, A\&AS, 134, 103
\reference{} Cassisi, S., \& Salaris, M. 1997, MNRAS, 285, 593
\reference{} Castellani, M., Castellani, V., Pulone, L., \& Tornamb\'e, A. 1994,
A\&A, 282, 711
\reference{} Castellani, V., Degl'Innocenti, S., \& Luridiana, V.  1993, A\&A, 
272, 442
\reference{} Castellani, M., \& Tornamb\'e, A. 1991, ApJ, 381, 393
\reference{} Copi, C. J., Olive, K. A., \& Schramm, D. N. 1995, ApJ, 451, 51
\reference{} Copi, C. J., Olive, K. A., \& Schramm, D. N. 1996, astro-ph/9606156
\reference{} de Bernardis, P., et al. 2000, Nature, 404, 955
\reference{} de Bernardis, P., et al. 2001, ApJ, submitted, astro-ph/0105296
\reference{} De Santis, R., \& Cassisi, S. 1999, MNRAS, 308, 97
\reference{} Di Bari, P. \& Foot, R. 2001, Phys. Rev. D, 63, 043008
\reference{} D'Odorico, S., Dessauges-Zavadsky, M., \& Molaro, P. 2001, 
A\&A, 368, L21
\reference{} Dorman, B., Rood, R. T., \& O'Connell, R. W. 1993, ApJ, 419, 596
\reference{} Dolgov, A. D. 1992, Phys. Rep., 222, 6  
\reference{} Dolgov, A. D., \& Pagel, B. E. J. 1999, NewA, 4, 223 (DP) 
\reference{} Esposito, S., et al., 2001, Phys. Rev. D, 63, 043004
\reference{} Ferraro, F. R., Montegriffo, P., Origlia, L., \& Fusi Pecci, F. 
2000, AJ, 119, 1282
\reference{} Frogel, J. A., Persson, S. E., \& Cohen, J. G. 1983, ApJS, 53, 713
\reference{} Giannone, P., \& Rossi, L. 1981, A\&A, 102, 386 
\reference{} Greggio, L., \& Renzini, A. 1990, ApJ, 364, 35 
\reference{} Greggio, L., \& Renzini, A. 1999, Mem. Soc. Astr. It., 70, 691 
\reference{} Gruenwald, R., Steigman, G. \& Viegas, S. M. 2000, ApJ, accepted,  
astro-ph/0109071 
\reference{} Halverson, N.W. et al. 2001, ApJ, submitted, astro-ph/0104489
\reference{} Hanany, S., et al., 2000, ApJL, 545, L5
\reference{} Hu, W. \& White, M. 1996, ApJ, 471, 30
\reference{} Iben, I. Jr. 1968, Nature, 220, 143 
\reference{} Izotov, Y. I., \& Thuan, T. X. 1998, ApJ, 497, 227
\reference{} Izotov, Y. I., Thuan, T. X., \& Lipovetsky, V. A. 1997, ApJS, 108, 1
\reference{} Jaffe, A., et al. 2001, Phys. Rev. Lett. 86, 3475
\reference{} Jungman, G., et al. 1996, Phys.Rev. D54, 1332
\reference{} Kamionkowski, M. \& Kosowsky, A. 1999, Ann.Rev.Nucl.Part.Sci. 49, 77-123
\reference{} Kurki-Suonio, H. \& Sihvola, E., 2001, Phys. Rev. D, 63, 083508
\reference{} Lange, A. et al., 2001, Phys. Rev. D, 63, 042001
\reference{} Lesgourgues, J. \& Peloso, M. 2000, Phys. Rev. D, 62, 081301
\reference{} Lee, M. G., Freedman, W. L., \& Madore, B. F. 1993, ApJ, 417, 533
\reference{} Lee, A.T., et al. 2001, ApJ, 561, L1  
\reference{} Mandolesi, N., et al. 1998, {\sc Planck} Low Frequency Instrument, 
a proposal submitted to ESA 
\reference{} Moehler, S., Sweigart, A.\ V., Landsman, W.\ B., Heber, U.,
\& Catelan, M.  1999, \aap, 346, L1 
\reference{} Netterfield, B., et al. 2001, ApJ, submitted, astro-ph/0104460 
\reference{} Olive, K. A., \& Steigman, G. 1995, ApJS, 97, 49 
\reference{} Olive, K. A., Steigman, G., \& Skillman, E. D. 1997, ApJ, 483, 788 
\reference{} Olive, K. A., Steigman, G., \& Walker, T. P. 2000, 
Phys. Rep., 333, 389  
\reference{} \"Ostlin, G. 2000, ApJ, 535, L99 
\reference{} Padin, S., et al. 2001, ApJ, 549, L1
\reference{} Peimbert, M., Peimbert, A., \& Luridiana, V. 2001, ApJ, accepted,
astro-ph/0107189  
\reference{} Pistinner, S. L., Hauschildt, P. H., Eichler, D., \& Baron, E. A. 
1999, MNRAS, 302, 684
\reference{} Pryke, C., et al., 2001, ApJ, submitted, astro-ph/0104490
\reference{} Puget, J. L., et al. 1998, {\sc Planck} High Frequency Instrument, 
a proposal submitted to ESA 
\reference{} Raffelt, G. G. 1990, ApJ, 365, 559
\reference{} Salaris, M., \& Cassisi, S. 1997, MNRAS, 289, 406
\reference{} Salaris, M., \& Cassisi, S. 1998, MNRAS, 298, 166
\reference{} Salaris, M., Chieffi, A., \& Straniero, O. 1993, ApJ, 414, 580
\reference{} Sandquist, E. L. 2000, MNRAS, 313, 571
\reference{} Sasselov, D., \& Goldwirth, D. 1995, ApJ, 444, L5
\reference{} Sauer, D., \& Jedamzik, K. 2001, A\&A, submitted, astro-ph/0104392 
\reference{} Stompor, R. et al. 2001, ApJ, 561, L7  
\reference{} Sweigart, A. V., \& Gross, P. G. 1976, ApJS, 32, 367
\reference{} Sweigart, A. V., \& Gross, P. G. 1978, ApJS, 36, 405
\reference{} Tantalo, R., Chiosi, C., Bressan, A., \& Fagotto, F. 1996, 
A\&A, 311, 361  
\reference{} Tegmark, M., \& Zaldarriaga, M. 2000, ApJ, 544, 30
\reference{} Tegmark, M., Zaldarriaga, M., \& Hamilton A. J. S. 2001, 
Phys. Rev. D., 63, 043007  
\reference{} Vandenberg, D. A., Stetson, P. B., \& Bolte, M. 1996, 
ARA\&A, 34, 461
\reference{} VandenBerg, D. A., Swenson, F. J., Rogers, F. J.,, Iglesias, C. A., \& Alexander, D. R. 2000, ApJ, 532, 430
\reference{} Viegas, S. M., Gruenwald, R., \& Steigman, G. 2000, ApJ, 531, 813
\reference{} Wright, E.L., 1999, New Astr.Rev.,  43, 257 
\reference{} Yi, S., Lee, Y.-W., Woo, J.-H., Park, J.-H., Demarque, P., 
Oemler, A. Jr. 1999, ApJ, 513, 128  
\reference{} Zoccali, M., Cassisi, S., Piotto G., Bono G., \& Salaris M.
1999, ApJ, 518, L49 
\reference{} Zoccali, M., Cassisi, S., Bono, G., Piotto, G., Rich, R.  M., 
\& Djorgovski, S. G. 2000, ApJ, 538, 289  
\end{references}
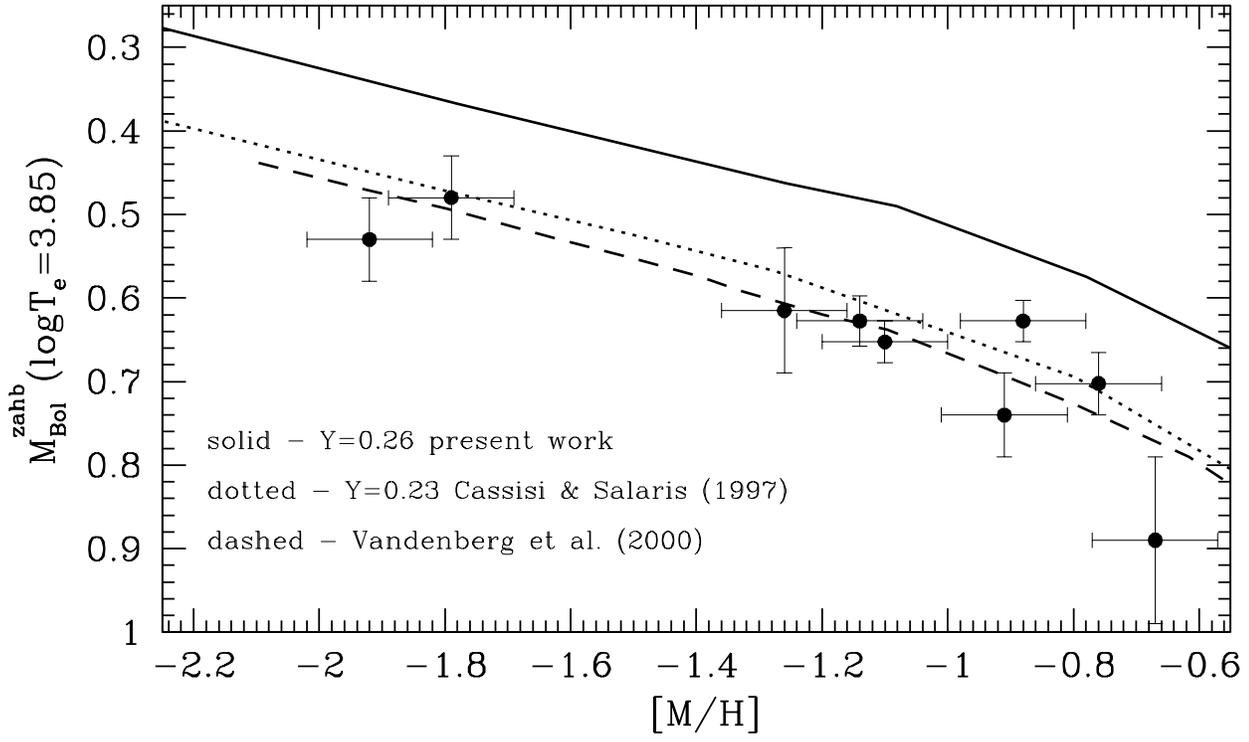
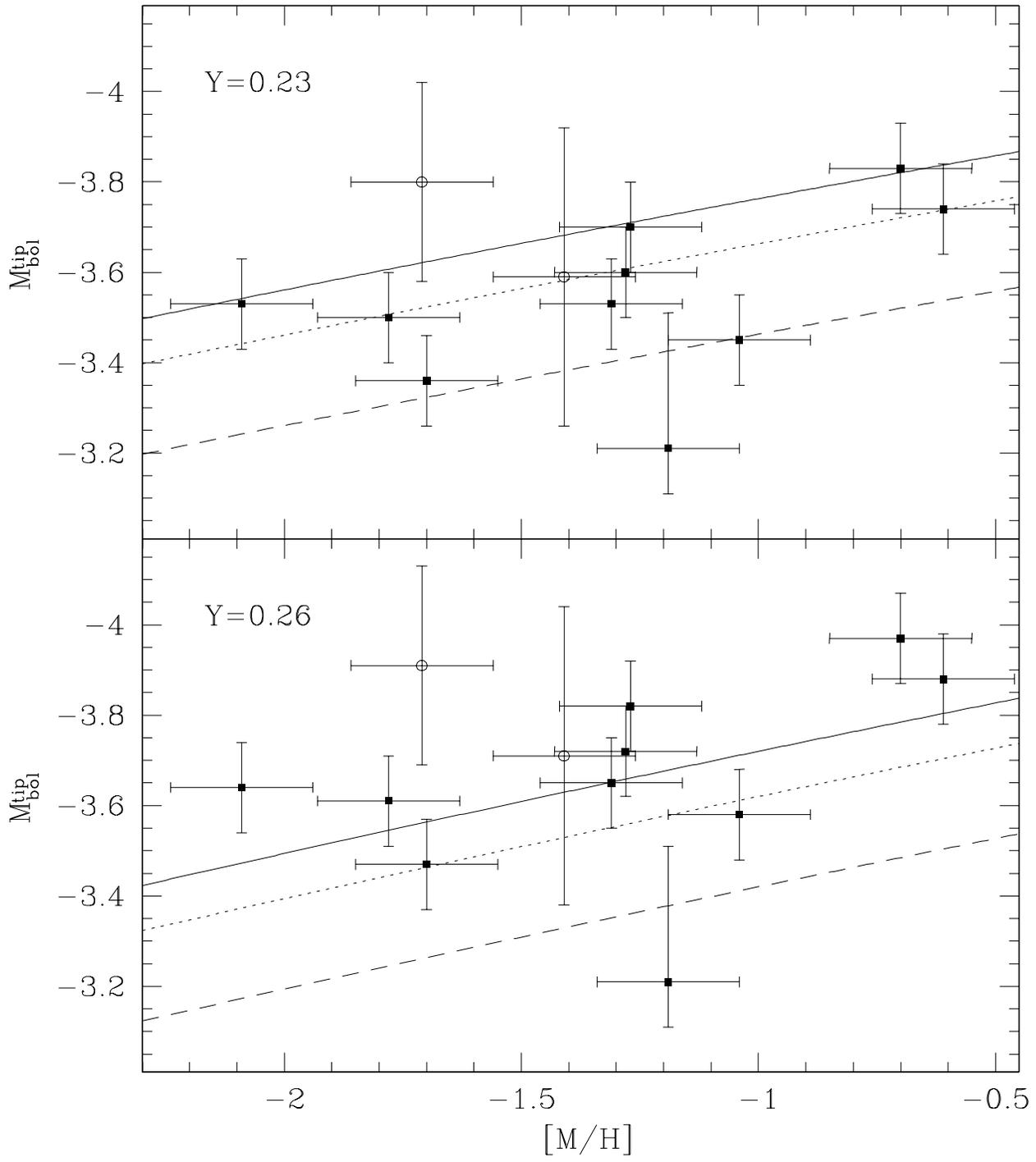
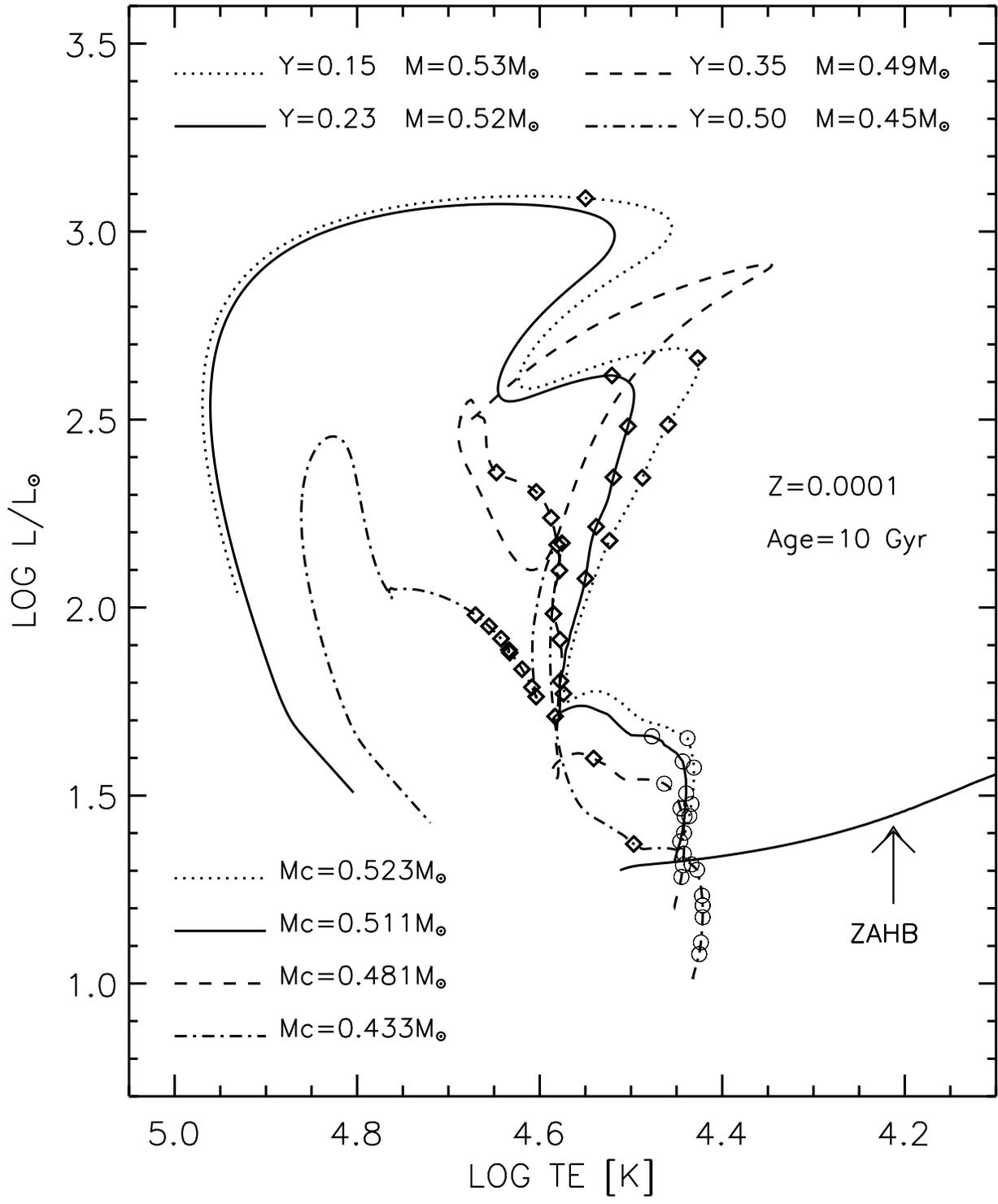
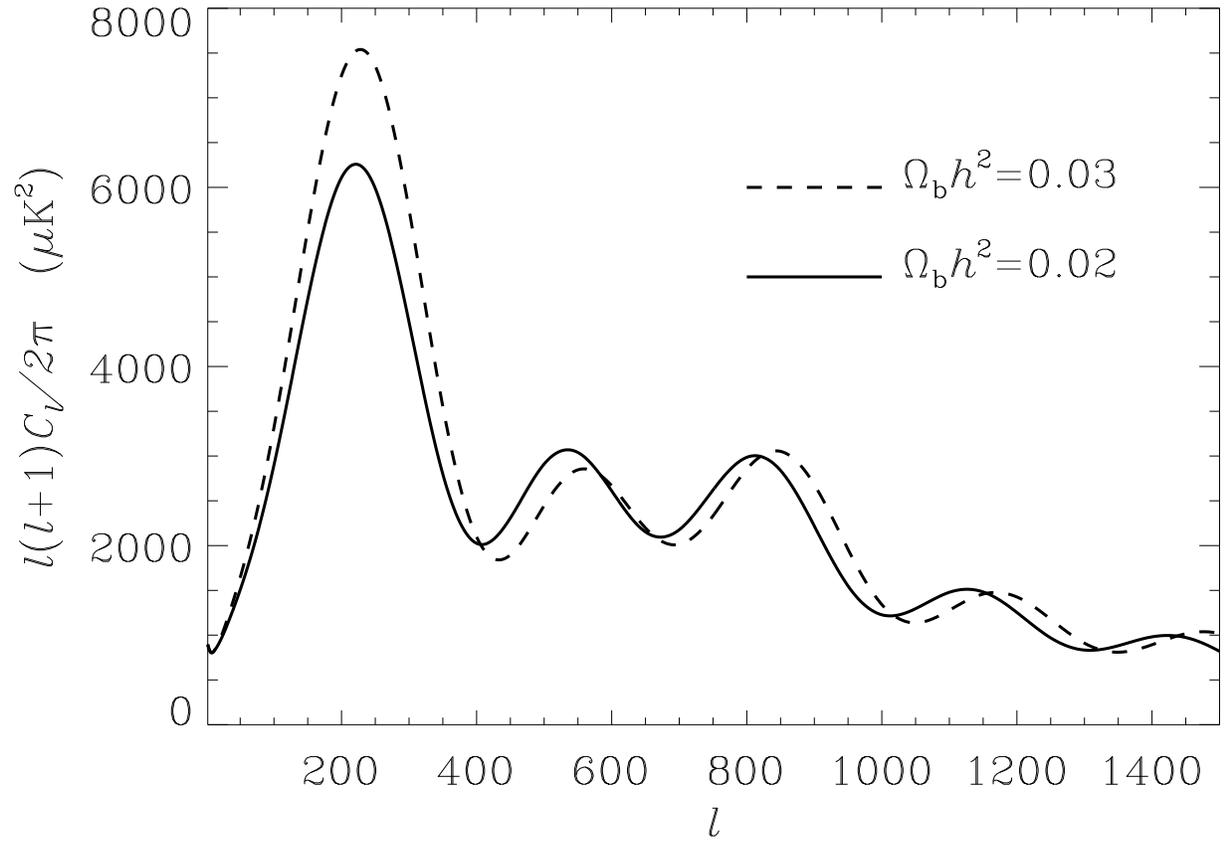
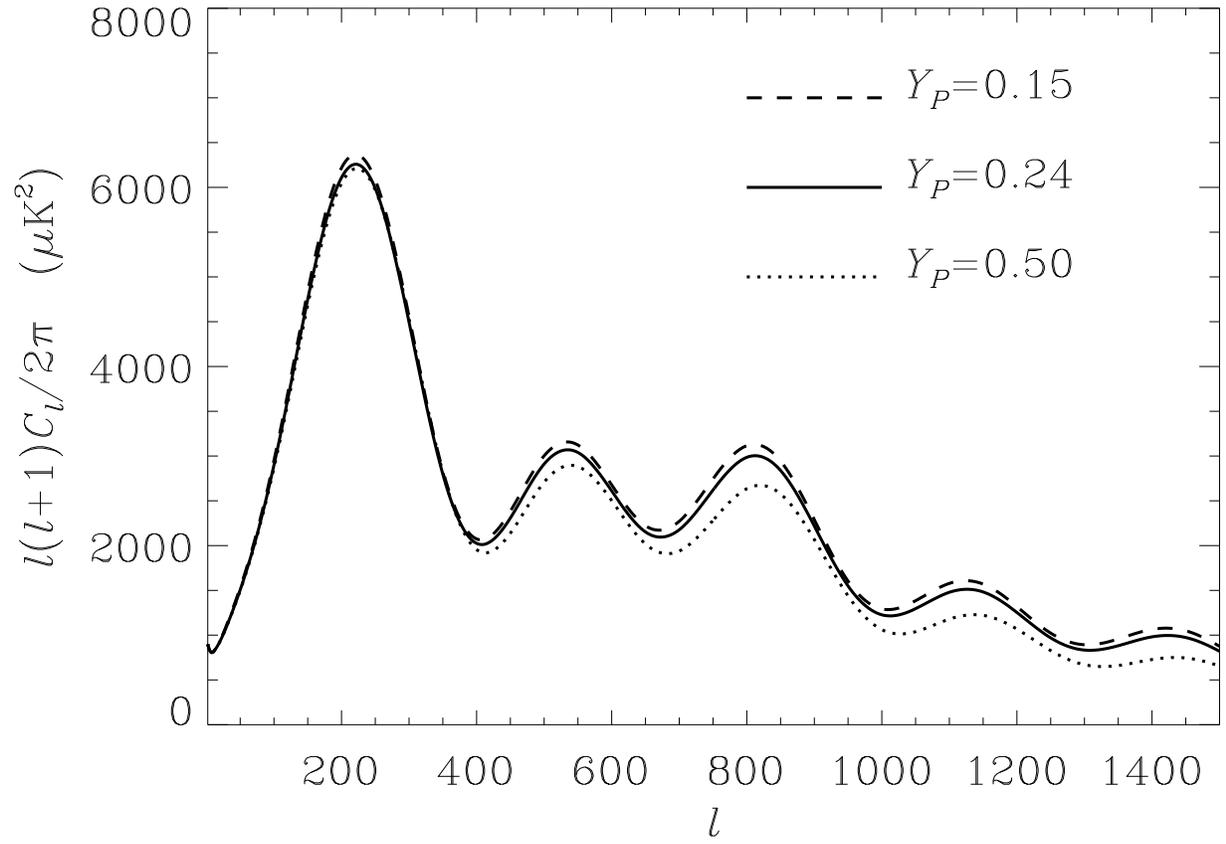
\end{document}